\documentclass[preprint,review,12pt]{elsarticle}
\usepackage[a4paper,margin=1in]{geometry}

\makeatletter
\def\ps@pprintTitle{%
\let\@oddhead\@empty
\let\@evenhead\@empty
\def\@oddfoot{}%
\let\@evenfoot\@oddfoot}
\makeatother
\usepackage{graphicx, subcaption}
\usepackage{epstopdf}
\usepackage{color}
\usepackage{multirow}
\usepackage{textcomp,gensymb}
\usepackage{xcolor}
\usepackage{physics}
\usepackage{epsfig}
\usepackage{amsmath}
\usepackage{amssymb}
\usepackage{booktabs}
\usepackage{siunitx}
\usepackage{breqn}
\biboptions{square,comma,sort&compress}
\usepackage{hyperref}
\usepackage{float}
\usepackage{subfiles}
\journal{}
\begin{document}

\begin{frontmatter}

\title{First-Principles-Informed Modeling of Thermal Transport and Laser-Induced Heating in Ultrahigh-Temperature Carbides} 

\author[1]{Shivanand Yadav}
\author[1]{Shailendra Kumar Verma}
\author[1]{Vardhman Dwivedi}
\author[1]{Raihan Ahammed}
\author[3]{Rohitash Kumar}
\author[3]{Manvendra Sharma}
\author[4]{Pritam Chakraborty}
\author[1]{Somnath Bhowmick}
\ead{bsomnath@iitk.ac.in}

\affiliation[1]{Department of Materials Science and Engineering, Indian Institute of Technology Kanpur, UP 208016, India}
\affiliation[3]{Defence Laboratory Jodhpur, Rajasthan 342011, India}
\affiliation[4]{Department of Aerospace Engineering, Indian Institute of Technology Kanpur, UP 208016, India}

\begin{abstract}
Using first-principles density functional theory and Boltzmann transport calculations, we systematically investigate the electronic structure, lattice and electronic thermal transport, optical response, and laser-induced heating of ultrahigh-temperature carbides (UHTCs), including Group IV (TiC, ZrC, and HfC) and Group V carbides (VC, NbC, and TaC), and SiC. The transition-metal carbides (TMCs) exhibit metallic electronic structures, whereas SiC exhibits semiconducting character. The calculated thermal transport properties reveal distinct contributions from phonons and electrons, with electronic heat transport becoming increasingly important at elevated temperatures in TMCs. The Group V carbides exhibit substantially lower lattice thermal conductivity than their Group IV counterparts, accompanied by pronounced Kohn anomalies and modified phonon dispersions associated with their distinct electronic structures, suggesting a possible connection between Fermi-surface topology, phonon dynamics, and lattice thermal transport. In contrast, SiC exhibits comparatively high lattice thermal conductivity, consistent with its long-lived acoustic phonons, but no electronic contribution due to its semiconducting nature. Optical calculations reveal relatively high reflectivity in the visible and infrared spectral ranges for the metallic carbides, whereas SiC exhibits lower reflectivity in the infrared region. Finally, one-dimensional and three-dimensional continuum heat-conduction models are employed to compare the thermal responses of these materials under specified laser irradiation conditions. The predicted temperature distributions depend on the thermal and optical properties of the materials as well as the specimen geometry. Although the adopted thermal model does not explicitly account for phase changes or latent heat, the results provide a first-principles-informed basis for comparing laser-induced heating and screening candidate materials for laser-resistant applications.
\end{abstract}

\end{frontmatter}

\section{Introduction}
Managing extreme thermal loads is a central challenge~\cite{QIU19942789} in advanced aerospace, defense, energy, and manufacturing technologies. Materials used in hypersonic vehicles, thermal protection systems, laser-resistant coatings, and nuclear environments must withstand severe thermal conditions while maintaining their structural integrity. Ultrahigh-temperature carbides (UHTCs), including transition-metal carbides and silicon carbide, are promising candidates for such applications because of their high melting temperatures, mechanical robustness, and chemical stability~\cite{savvatimskiy2017measurement, guillermet1995analysis, snead2010effects, katoh2013properties, protik2017phonon, samolyuk2011molecular, yeo2013enhanced, schwind2022thermal, SHIREMAN2026128878}. However, their response to intense thermal and radiative loading depends not only on their intrinsic thermal stability but also on their ability to absorb, conduct, and redistribute heat. In particular, the interplay among lattice vibrations, conduction electrons, and optical properties can influence the temperature distribution that develops during laser irradiation. Understanding these contributions is therefore important for assessing the thermal response of UHTCs and identifying materials with potential for laser-resistant and other extreme-environment applications.

Heat transport in carbides involves contributions from lattice vibrations (phonons) and, in metallic systems, conduction electrons. The relative contributions of these two mechanisms depend on the electronic structure, phonon dispersion, and temperature. In semiconducting SiC, heat transport is predominantly phonon-mediated, whereas both phononic and electronic contributions can be important in transition-metal carbides (TMCs). In particular, electronic thermal conductivity can make a substantial contribution to the total thermal conductivity of metallic TMCs, with its importance potentially increasing at elevated temperatures. A reliable assessment of their thermal transport therefore requires consideration of both lattice and electronic contributions. The lattice contribution is governed by phonon group velocities and lifetimes, which depend on interatomic bonding, phonon dispersion, and anharmonic phonon--phonon scattering. In metallic carbides, the electronic structure and Fermi-surface topology can also influence phonon dispersions through Kohn anomalies at selected wave vectors. Examining both electronic and lattice thermal transport is therefore essential for understanding the thermal behavior of these materials over a broad temperature range. 

Although the thermal properties of several ultrahigh-temperature carbides have been investigated experimentally and computationally, previous studies have often focused on individual compounds or specific physical properties. In particular, despite the potential importance of electronic heat transport in metallic transition-metal carbides, published information on their electronic thermal conductivity remains limited. To the best of our knowledge, ZrC is the only compound among the carbides considered in the present study for which electronic thermal conductivity has been reported in the literature~\cite{Tiwari23}. Comparable information for other Group IV and Group V carbides is therefore limited, making a systematic assessment of their electronic thermal transport valuable. Furthermore, a consistent comparison of electronic and lattice thermal conductivity across these carbide families, together with their phonon dynamics and optical response, can help clarify the factors governing their thermal behavior. Combining these properties with continuum heat-conduction modeling also provides a framework for comparing their responses under specified laser irradiation conditions~\cite{Khan25, Anirudh23, Xu26}. 

In this work, we investigate Group IV transition-metal carbides (TiC, ZrC, and HfC), Group V transition-metal carbides (VC, NbC, and TaC), and SiC using first-principles calculations. We examine their electronic structures, phonon dispersions, lattice and electronic thermal transport over a broad temperature range of 300--3500~K, and optical properties. Particular attention is given to quantifying the electronic and lattice contributions to thermal conductivity across the investigated carbide families, thereby enabling a consistent comparison of their relative importance. We also focus on the differences in phonon dynamics and thermal transport between Group IV and Group V carbides, including the possible relationship between Fermi-surface nesting, Kohn anomalies, and lattice thermal conductivity. The calculated thermal and optical properties are then incorporated into one-dimensional and three-dimensional continuum heat-conduction models to compare the temperature responses of these materials under specified laser irradiation conditions. By combining atomistic calculations with continuum thermal modeling, this study provides a comparative assessment of the factors governing heat transport and laser-induced heating in the investigated carbides, with implications for screening candidate materials for laser-resistant and other extreme-thermal applications.

\section{Methodology}
First-principles calculations are performed using density functional theory (DFT) and density-functional perturbation theory (DFPT), as implemented in the Quantum ESPRESSO package~\cite{giannozzi2009jpcm, giannozzi2017jpcm, baroni2001rmp}.  The lattice parameters are fixed at their experimental values. In transition-metal carbides, the metal and carbon atoms occupy the $4a$ and $4b$ Wyckoff positions, respectively, in the cubic rocksalt structure. Owing to the centrosymmetric crystal structure and the corresponding site symmetries, no internal structural relaxation is required. Optimized norm-conserving Vanderbilt pseudopotentials are employed~\cite{hamann2013prb,schlipf2015cpc} and the Kohn--Sham wavefunctions are expanded in a plane-wave basis with a kinetic energy cutoff of 100~Ry. Exchange--correlation effects are treated using the Perdew--Burke--Ernzerhof generalized-gradient approximation (PBE-GGA)~\cite{PhysRevLett.77.3865}. Fermi--Dirac smearing with a width of 0.01~Ry is used to describe the electronic occupations of the metallic systems. The self-consistent electronic calculations employ a $24\times24\times24$ Monkhorst--Pack $k$-point mesh~\cite{monkhorst1976prb}. The dense reciprocal-space sampling and smearing parameters are selected to resolve the electronic states near the Fermi level, which are important for describing the Fermi-surface features and associated Kohn anomalies in the Group~V carbides.

\begin{figure*}
    \centering{\includegraphics[width=1.00\linewidth]{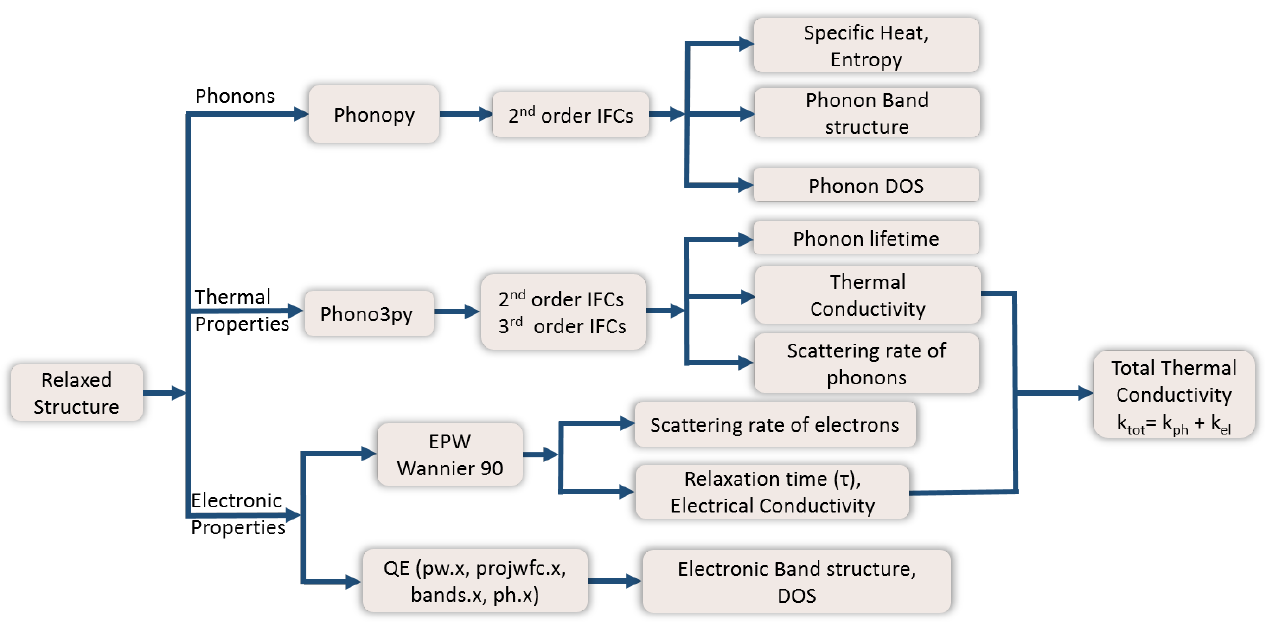}}
     \caption{Schematic representation of the DFT-based computational workflow used in this study to obtain the thermodynamic and transport properties.}
    \label{fig1}
\end{figure*}

The dynamical matrices are calculated using density-functional perturbation theory (DFPT) on an $8\times8\times8$ $q$-point grid, with a self-consistency threshold of $10^{-14}$. The resulting dynamical matrices are Fourier transformed to obtain the real-space interatomic force constants. For SiC, the non-analytic correction is included using the Born effective charges and electronic dielectric tensor to account for the longitudinal--transverse optical (LO--TO) phonon splitting at the Brillouin-zone center ($\Gamma$). Electron--phonon interactions and electronic transport coefficients are calculated using the EPW package within a maximally localized Wannier function framework~\cite{PhysRevB.76.165108, PhysRevB.56.12847, PhysRevB.65.035109}. This approach provides the electronic transport properties required to evaluate the electronic contribution to thermal conductivity, enabling a systematic comparison of electronic and lattice heat transport across the investigated transition-metal carbides. Figure \ref{fig1} summarizes the computational workflow adopted in this study, from the first-principles electronic-structure and lattice-dynamics calculations to the evaluation of the electronic and lattice contributions to thermal conductivity. The computational details for the individual properties are described in the respective subsections, Appendix, and Supplementary Material.

\section{Results and discussion}
\subsection{Crystal and Electronic Band Structure}
\begin{figure*}
    \centering{\includegraphics[width=\linewidth]{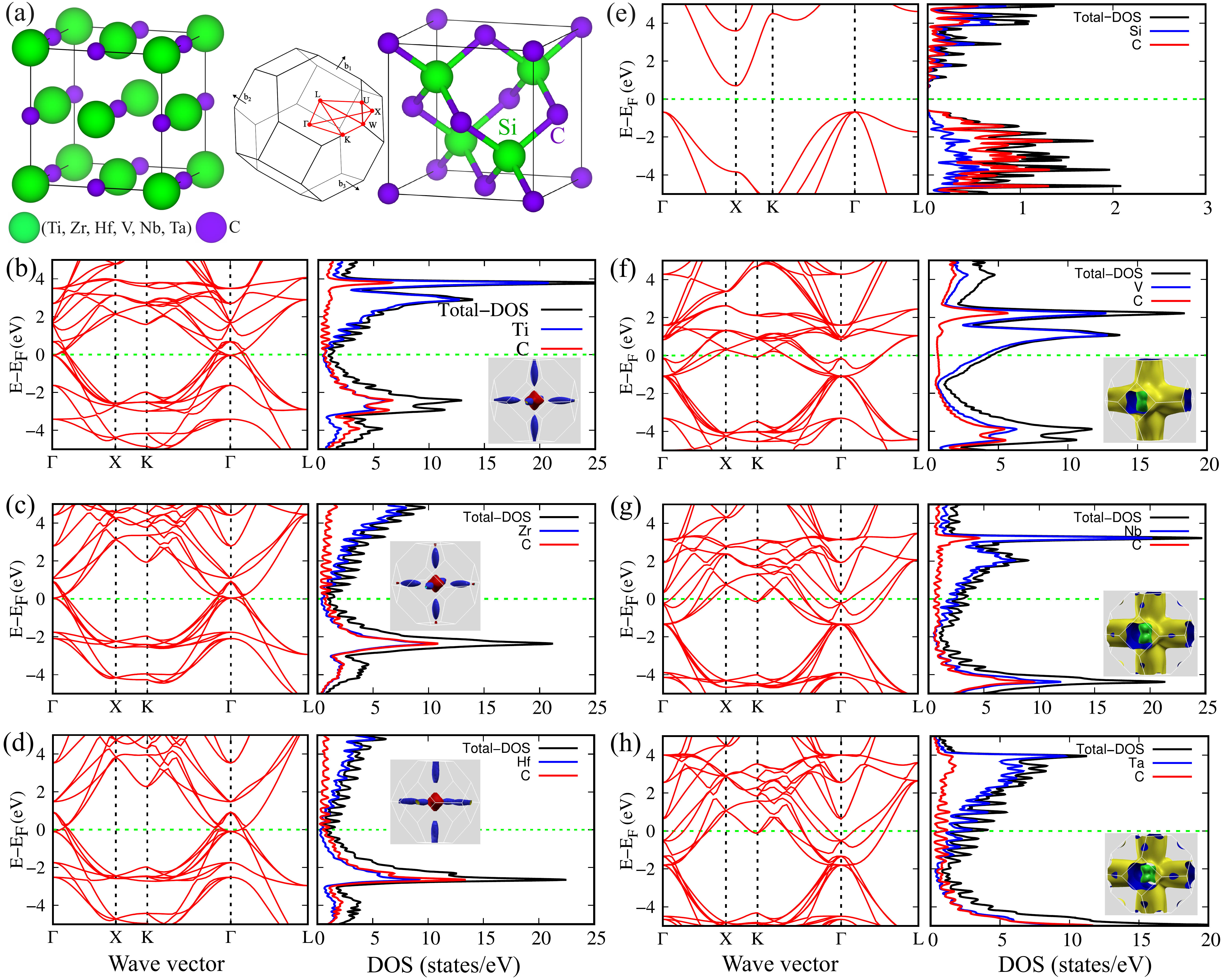}}
     \caption{ Crystal structure of UHTCs and the corresponding Brillouin zone, showing the high-symmetry points. Electronic band structure and corresponding atom-projected density of states (DOS) of (b--d) TiC, ZrC, and HfC, and (e--h) SiC, VC, NbC, and TaC. The inset illustrates the Fermi surface plots. The Fermi surfaces in panels (b--d) exhibit no significant Fermi nesting, whereas clear Fermi surface nesting features are observed in panels (f--h).}
    \label{fig2}
\end{figure*}

The crystal structures of the transition-metal carbides $MC$ ($M=\mathrm{Ti}$, $\mathrm{Zr}$, $\mathrm{Hf}$, $\mathrm{V}$, $\mathrm{Nb}$, and $\mathrm{Ta}$) and SiC are shown in Figure~\ref{fig2}(a), together with the corresponding Brillouin zones and high-symmetry paths used for the electronic-structure and lattice-dynamics calculations. The transition-metal carbides considered here adopt the cubic rocksalt structure. The crystal structure of SiC is distinct and is specified separately in Figure~\ref{fig2}(a). These structural representations provide the basis for comparing the calculated electronic structures and transport properties of the investigated materials. 

The electronic band structures of the investigated materials along the selected high-symmetry directions are shown in Figure~\ref{fig2}(b--h). The transition-metal carbides exhibit several bands crossing the Fermi level ($E_{\mathrm{F}}$), indicating their metallic character. The bands near $E_{\mathrm{F}}$ display appreciable dispersion, reflecting the variation of electronic energies with wave vector and the associated band velocities. These electronic states can contribute to charge and heat transport. However, the magnitude of the electronic thermal conductivity depends not only on the band dispersion but also on the electronic scattering rates and carrier lifetimes, which are considered in the subsequent transport calculations.

In contrast to the metallic transition-metal carbides, SiC exhibits semiconducting behavior, with an indirect band gap of 1.37~eV. The valence-band maximum is located at the $\Gamma$ point, whereas the conduction-band minimum occurs near the $X$ point, as shown in Figure~\ref{fig2}(e). The calculated band gap is smaller than the experimental value of 2.36~eV~\cite{levinshtein2001properties}, consistent with the well-known tendency of semilocal exchange--correlation functionals to underestimate semiconductor band gaps. Despite this quantitative difference, the calculated band structure captures the semiconducting character of SiC, distinguishing its electronic structure from those of the metallic transition-metal carbides.

The atom-projected density of states (PDOS) provides further insight into the orbital contributions to the electronic structures. For the transition-metal carbides, the states near the Fermi level are dominated by transition-metal $d$ orbitals, with a smaller but non-negligible contribution from carbon $p$ states. This orbital character indicates appreciable metal--carbon hybridization, consistent with the mixed metallic and covalent character of bonding in these compounds. The electronic states near the Fermi level are particularly relevant to electronic transport, although their contribution to the transport coefficients also depends on their dispersion and scattering rates.

For SiC, the valence-band edge is dominated by carbon $p$ states, whereas the conduction-band edge has a substantial contribution from silicon $ p$ orbitals. These orbital projections are consistent with the distinct electronic structure and bonding character of SiC relative to those of metallic transition-metal carbides.

A comparison of the Group~IV carbides (TiC, ZrC, and HfC) with the Group~V carbides (VC, NbC, and TaC) reveals differences in their electronic structures near the Fermi level. Although both families exhibit metallic behavior, the Group~V carbides show a larger density of states near $E_{\mathrm{F}}$ in the present calculations. This difference is consistent with a greater contribution from partially occupied transition-metal $d$ states in the Group~V compounds. Such differences in the electronic structure may influence electron--phonon interactions and electronic transport. The Fermi-surface geometries provide additional information about the momentum-dependent electronic response and are examined below in relation to the observed differences between the two carbide families.

Fermi-surface nesting provides a useful geometrical perspective on the momentum-dependent electronic response to lattice vibrations~\cite{PhysRevB.8.5082}. As shown in the insets of Figure~\ref{fig2}(b-d), the Fermi surfaces of TiC, ZrC, and HfC are predominantly characterized by curved and compact sheets, with no prominent extended, nearly parallel regions. This geometry suggests relatively weak nesting compared with a Fermi surface containing large, approximately parallel sections. Consequently, the Fermi-surface geometry does not indicate extensive connection of different regions of the Fermi surface by a single scattering vector. However, the Fermi-surface geometry alone does not determine the electronic response and electron--phonon scattering; they also depend on the relevant electronic states and coupling matrix elements. 

In contrast, the Fermi surfaces of VC, NbC, and TaC exhibit more extended regions, including approximately parallel portions, as illustrated in the insets of Figure~\ref{fig2}(f-h). These features are consistent with more pronounced Fermi-surface nesting, whereby a characteristic wave vector can connect extended regions of the Fermi surface. Such nesting can enhance the momentum-dependent electronic response to lattice distortions and may contribute to the development of Kohn anomalies in the phonon dispersion~\cite{PhysRevB.28.977, PhysRevLett.29.353, PhysRevB.77.180507}. The relationship between these Fermi-surface features and the calculated phonon dispersions is discussed in the following section.

The differences in Fermi-surface (FS) nesting between the two carbide groups are associated with their distinct electronic structures. Compared with the Group~IV carbides, the Group~V transition-metal carbides have one additional valence electron per formula unit, which modifies the occupation of the metal--carbon hybridized states near the Fermi level. The resulting changes in band filling are accompanied by differences in the FS geometry. In TiC, ZrC, and HfC, the FSs are predominantly curved, with relatively few extended, nearly parallel regions. In contrast, VC, NbC, and TaC exhibit more extended, locally parallel FS regions, consistent with more pronounced nesting. This difference provides a qualitative electronic-structure basis for understanding the distinct momentum-dependent electronic responses of the two carbide groups and may be relevant to the differences in their phonon dispersions and associated Kohn anomalies, discussed in the following section.

\begin{figure*}
\centering
\includegraphics[width=\linewidth]{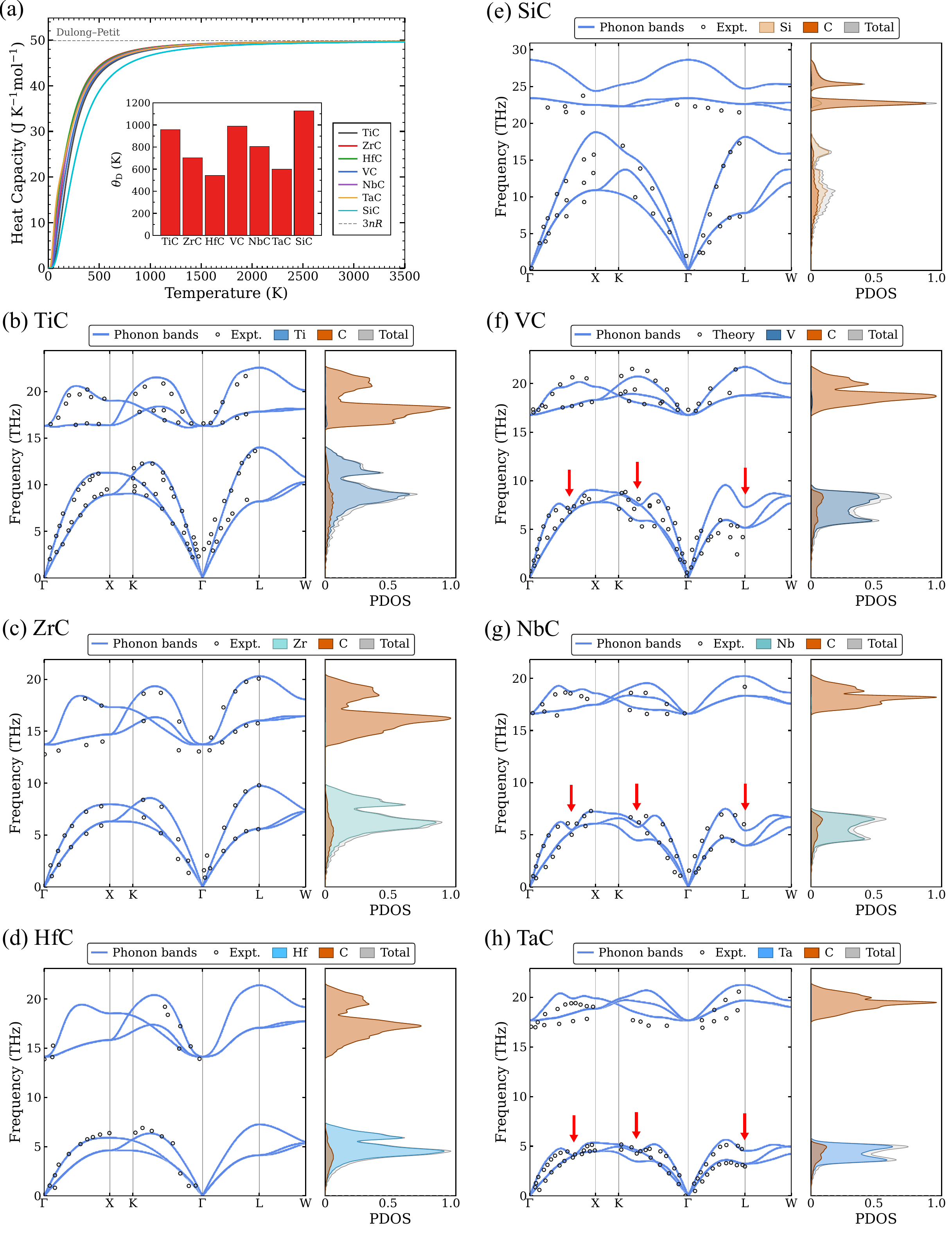}
\caption{(a) Heat capacity as a function of temperature. The inset shows the Debye temperatures obtained from Eq.~\ref{eq:dt}. (b--h) Phonon band structure and corresponding projected density of states. Black circles represent experimentally reported phonon frequencies~\cite{weber1973lattice, pintschovius1978lattice, smith1976phonon, smith1970phonon, serrano2002determination}; and theoretically reported frequencies for VC~\cite{isaev2007phonon}. Red arrows in (f--h) mark Kohn anomalies along $\Gamma$--X ($[\zeta00]$), K--$\Gamma$ ($[\zeta\zeta0]$), and $\Gamma$--L ($[\zeta\zeta\zeta]$). The group IV TMCs (b--d) and semiconducting SiC (e) show acoustic branches that vary smoothly across the zone.}
\label{fig3}
\end{figure*}

\subsection{Heat Capacity, Debye Temperature, and Phonon Band Structure}
Figure~\ref{fig3}(a) shows the temperature dependence of the constant-volume heat capacity, $C_V$, of the Group~IV
transition-metal carbides, Group~V transition-metal carbides, and SiC. The heat capacity is calculated by summing the
contributions of the phonon modes over the Brillouin zone using the interpolated phonon spectrum~\cite{baroni2001phonons}:
\begin{equation}
C_V(T)=\sum_{\mathbf{q}\nu}
k_{\mathrm B}
\frac{x_{\mathbf{q}\nu}^{2}}
{4\sinh^{2}(x_{\mathbf{q}\nu}/2)},
\qquad
x_{\mathbf{q}\nu}=
\frac{\hbar\omega_{\mathbf{q}\nu}}{k_{\mathrm B}T}.
\end{equation}
The summation is performed on a uniform $24\times24\times24$ $q$-point mesh. At low temperatures, only a limited portion of the phonon spectrum is thermally populated, resulting in a relatively small heat capacity. As the temperature increases, progressively more vibrational modes contribute, and the heat capacity rises. At sufficiently high temperatures, the calculated values approach the Dulong--Petit limit of $3nR$, where $n$ is the number of atoms per formula unit. For the binary compounds considered here, this limit is approximately $50~\mathrm{J\,mol^{-1}\,K^{-1}}$, consistent with the
high-temperature behavior shown in Figure~\ref{fig3}(a). The temperature-dependent thermodynamic properties are shown in Section S1 and Figure S1 of the Supplementary Material.

The Debye temperature, $\Theta_{\mathrm D}$, provides a characteristic energy scale for lattice vibrations and can be estimated from the average sound velocity, $v_m$, as
\begin{equation}
\label{eq:dt}
\Theta_{\mathrm{D}}=
\frac{\hbar}{k_{\mathrm{B}}}
\left[
\frac{6\pi^{2}n}{V_{0}}
\right]^{\frac{1}{3}}
v_{m},
\end{equation}
where $\hbar$ is the reduced Planck constant, $k_{B}$ is the Boltzmann constant, $n$ is the number of atoms in the unit cell and $V_0$ is its volume. Under the isotropic approximation, the average
sound velocity is given by
\begin{equation}
\label{eqvm}
v_{m}=
\left[
\frac{1}{3}
\left(
\frac{2}{v_{t}^{3}}
+
\frac{1}{v_{l}^{3}}
\right)
\right]^{-\frac{1}{3}}.
\end{equation}
where $v_t$ and $v_l$ are the transverse and longitudinal sound velocities, respectively. This expression assumes two degenerate transverse branches and direction-independent sound velocities. 

Because these assumptions are only approximate for cubic crystals, we also evaluate the Debye average using the direction-dependent acoustic velocities in its general form,
\begin{equation}
\label{eqvm1}
\frac{1}{v_{m}^{3}}
=\frac{1}{3}
\left\langle
\sum_{\nu=1}^{3}\frac{1}{v_{\nu}(\hat{\mathbf n})^{3}}
\right\rangle_{\hat{\mathbf n}}
=\frac{1}{12\pi}
\int
\sum_{\nu=1}^{3}\frac{d\Omega}{v_{\nu}(\hat{\mathbf n})^{3}},
\end{equation}
where $v_{\nu}(\hat{\mathbf n})$ is the velocity of acoustic branch $\nu$ propagating along $\hat{\mathbf n}$ and $\langle\cdots\rangle$ denotes an average over the unit sphere. Eq.~\ref{eqvm1} reduces to Eq.~\ref{eqvm} when $v_{1}=v_{2}=v_{t}$ and $v_{3}=v_{l}$ in every direction. The velocities are obtained directly from the slopes of the acoustic branches near the $\Gamma$ point along 300 directions distributed over the unit sphere. Although the individual acoustic velocities exhibit directional anisotropy, the resulting directionally averaged sound velocity differs by less than 2\% from the isotropic estimate for the materials considered. This comparison indicates that the isotropic approximation provides a reasonable estimate of the average sound velocity and the associated Debye temperature for the present dataset.

The Debye temperatures calculated from the average sound velocities for the Group~IV carbides, Group~V carbides, and SiC are presented in the inset of Figure~\ref{fig3}(a) and summarized in Table~\ref{tab:debye_ratio}.  The calculated values are generally consistent with previously reported results~\cite{liu2014first,  krasnenko2012first, herrero2025pressure, kostenko2022structural}. This agreement supports the reliability of the calculated long-wavelength acoustic velocities and the corresponding Debye-temperature estimates.


The phonon band structures of TiC, ZrC, HfC, and SiC along the selected high-symmetry directions are shown in Figure~\ref{fig3}(b–e), together with the corresponding phonon densities of states. No imaginary phonon frequencies are observed in the calculated dispersions, indicating dynamical stability of the considered structures within the harmonic approximation. The calculated phonon frequencies show agreement with the available experimental data, as illustrated by the comparison in Figure~\ref{fig3}(b–e).

The three acoustic branches, comprising one longitudinal acoustic (LA) mode and two transverse acoustic (TA) modes, exhibit approximately linear dispersion near the $\Gamma$ point, consistent with the long-wavelength elastic response of the lattice. The optical branches arise primarily from relative atomic vibrations within the unit cell. The phonon dispersions exhibit a separation between the acoustic and optical branches, while the projected phonon densities of states indicate that the low-frequency acoustic modes have a substantial metal-atom contribution and that carbon motion contributes strongly to the higher-frequency optical modes. The separation between these frequency ranges reflects the combined influence of atomic masses and interatomic force constants.

The phonon band structures of VC, NbC, and TaC are shown in Figure~\ref{fig3}(f--h). As in the Group~IV carbides, the dispersions exhibit distinct acoustic and optical frequency ranges. However, the acoustic branches of these Group~V carbides display pronounced dips along selected high-symmetry directions, including $\Gamma$--X ($[\zeta00]$), K--$\Gamma$ ($[\zeta\zeta0]$), and $\Gamma$--L ($[\zeta\zeta\zeta]$), as indicated by the red arrows in Figure~\ref{fig3}(f--h). These features are consistent with Kohn anomalies, which arise from the wave-vector-dependent electronic response to lattice distortions and the associated renormalization of phonon frequencies~\cite{PhysRevLett.2.393}.

Such anomalies can become pronounced at wave vectors that connect relevant regions of the Fermi surface, particularly when the electronic response varies strongly with wave vector. The more extended, approximately parallel Fermi-surface regions observed for VC, NbC, and TaC may therefore be relevant to the anomalies identified in their phonon dispersions. In contrast, the calculated acoustic branches of TiC, ZrC, HfC, and SiC vary more smoothly along the corresponding paths, without similarly pronounced dips. These results indicate a difference in the momentum-dependent phonon response between the two carbide groups.

The Debye temperature obtained from the average sound velocity primarily reflects the long-wavelength acoustic response of the lattice and does not directly capture phonon softening at finite wave vectors. To examine the implications of the calculated low-frequency phonon spectrum for the thermodynamic response, we also estimate a Debye temperature from the calculated heat capacity at $40~\mathrm{K}$ using the low-temperature Debye relation,
\begin{equation}
C_V(T)=\frac{12\pi^4}{5}nR
\left(\frac{T}{\Theta_{\mathrm D}}\right)^3,
\end{equation}
where $n$ is the number of atoms per formula unit and $R$ is the gas constant. This expression assumes a Debye-like
low-frequency spectrum with approximately linear acoustic dispersion. The two Debye temperatures would coincide if the
calculated low-frequency spectrum were adequately represented by this approximation.

\begin{table*}
  \centering
  \caption{Comparison of the Debye temperature obtained from the average sound velocity, Eq.~(2), with that required to reproduce the calculated heat capacity at 40~K through the Debye $T^{3}$ law. A ratio below unity measures the excess of low-frequency acoustic modes produced by the Kohn anomalies. $M_{\mathrm{TM}}$ is the mass of the transition-metal (or Si) atom.}
  \label{tab:debye_ratio}
  \setlength{\tabcolsep}{8pt}
  \begin{tabular}{@{}l l *{7}{c} @{}}
  \toprule
   & & {TiC} & {ZrC} & {HfC} & {VC} & {NbC} & {TaC} & {SiC} \\
  \midrule
  \multirow{2}{*}{$\Theta_{\mathrm D}$ (K)}
    & from $v_{m}$          & 957 & 703 & 544 & 989 & 805 & 600 & 1127 \\
    & from $C_{V}$ at 40~K  & 861 & 565 & 413 & 703 & 508 & 380 & 1040 \\
  \addlinespace
  \multicolumn{2}{@{}l}{Ratio}
                            & 0.90 & 0.80 & 0.76 & 0.71 & 0.63 & 0.63 & 0.92 \\
  \multicolumn{2}{@{}l}{$M_{\mathrm{TM}}$ (amu)}
                            & 47.87 & 91.22 & 178.49 & 50.94 & 92.91 & 180.95 & 28.09 \\
  \bottomrule
  \end{tabular}
\end{table*}

As summarized in Table~\ref{tab:debye_ratio}, the ratio of the heat-capacity-based Debye temperature to that obtained from the average sound velocity is $0.90$, $0.80$, and $0.76$ for TiC, ZrC, and HfC, respectively, and $0.92$ for SiC. The corresponding ratios are lower for VC, NbC, and TaC, at $0.71$, $0.63$, and $0.63$, respectively. These differences indicate a greater deviation of the low-frequency vibrational spectrum from the linear-dispersion Debye approximation in the Group~V carbides. This trend is consistent with the pronounced softening of the acoustic branch identified in their calculated phonon dispersions. The comparison therefore provides a complementary thermodynamic perspective on the differences between the two carbide groups. However, the ratio reflects the overall low-frequency phonon spectrum and should not be interpreted as a definitive measure of Fermi-surface nesting or the strength of the Kohn anomaly.

\begin{figure*}
    \centering{\includegraphics[width=1\linewidth]{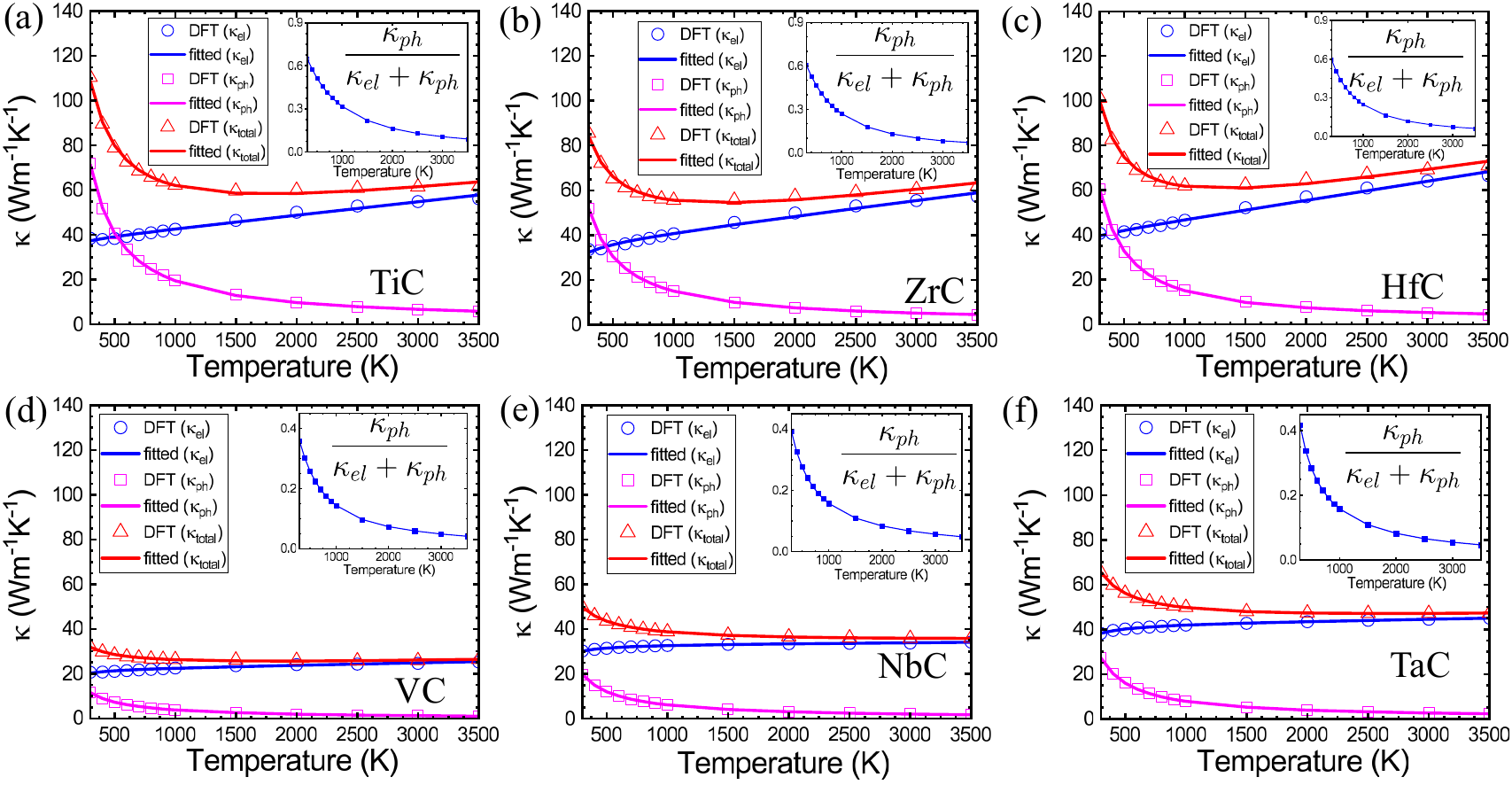}}
     \caption{Lattice (purple), electronic (blue), and total (red) thermal conductivity of group IV TMCs (TiC, ZrC, and HfC) in (a), (b), and (c), and of group V TMCs (VC, NbC, and TaC) in (d), (e), and (f). Points are calculated values, and lines are fitted. Inset: Ratio of phonon to total thermal conductivity. Near room temperature, total conductivity is dominated by phonons (electrons) in group IV (group V) TMCs. At high temperature, electronic conductivity heavily dominates in both group IV and V TMCs.}
    \label{fig5}
\end{figure*}

\subsection{Thermal Conductivity}
Except for SiC, the investigated materials are metallic transition-metal carbides in which both phonons and conduction electrons can contribute to heat transport. SiC, in contrast, exhibits semiconducting behavior, and its thermal conductivity is treated here in terms of the lattice contribution. The lattice and electronic thermal conductivities of the metallic carbides are calculated using first-principles-based transport methods. The computational procedures are described in the Methodology section and in \ref{app:ltc} and \ref{app:etc}, respectively.

The temperature dependence of the lattice thermal conductivity of the transition-metal carbides is shown in Figure~\ref{fig5}, while the corresponding results for SiC are provided in Figure~S2 of the Supplementary Material. For the investigated materials, the lattice thermal conductivity generally decreases with increasing temperature and exhibits an approximately inverse-temperature dependence over the range considered. This trend is primarily associated with the enhancement of anharmonic phonon--phonon scattering at elevated temperatures. The resulting reduction in phonon lifetimes and mean free paths suppresses lattice heat transport. This interpretation is consistent with the temperature-dependent, mode-resolved phonon scattering rates presented in Figures~S2--S4 of the Supplementary Material, which show an overall increase in phonon scattering with temperature.

The frequency-resolved lattice thermal conductivity presented in Figures~S2--S4 of the Supplementary Material indicates that acoustic phonons provide the dominant contribution to lattice heat transport. This result is consistent with the values summarized in Table~\ref{tab1}, which show that the acoustic contribution substantially exceeds the optical contribution for the investigated materials. Among the compounds considered, SiC exhibits the highest calculated lattice thermal conductivity, followed by the Group~IV and Group~V transition-metal carbides. The high lattice thermal conductivity of SiC is consistent with its relatively high phonon group velocities, which reflect the combined effects of its light constituent atoms and strong interatomic bonding. Phonon lifetimes, determined in part by anharmonic scattering, also influence the resulting thermal conductivity.

\begin{table*}
\centering
\caption{Acoustic phonon contribution ($\kappa_{ph}^{\mathrm{acoustic}}$), optical phonon contribution ($\kappa_{ph}^{\mathrm{optical}}$), total lattice thermal conductivity ($\kappa_{ph}^{\mathrm{total}}$), electronic thermal conductivity ($\kappa_{\mathrm{el}}$), and total thermal conductivity ($\kappa_{\mathrm{total}}$) at room temperature and 1000 K. Acoustic phonons dominate the lattice thermal conductivity part, while electrons are the major contributors at 1000 K.}
\resizebox{1.0\columnwidth}{!}{%
\begin{tabular}{c|ccccc|ccccc}
\hline
 & \multicolumn{5}{c|}{300 K} & \multicolumn{5}{c}{1000 K}  \\
\cline{2-11}
Materials 
& $\kappa_{ph}^{\mathrm{acoustic}}$  & $\kappa_{ph}^{\mathrm{optical}}$ & $\kappa_{ph}^{\mathrm{total}}$ & $\kappa_{ele}$ & $\kappa_{total}$ & $\kappa_{ph}^{\mathrm{acoustic}}$  & $\kappa_{ph}^{\mathrm{optical}}$ & $\kappa_{ph}^{\mathrm{total}}$ & $\kappa_{ele}$ & $\kappa_{total}$   \\
\hline
TiC  & 68.44 & 3.48 & 71.92 & 38.53 & 110.45  &  17.77 & 2.00 & 19.77 & 42.53 & 62.30  \\
ZrC  & 47.94 & 4.02 & 51.96 & 33.53 & 85.49 &  13.05 & 1.92 & 14.97 & 40.57 & 55.54   \\
HfC  & 58.11 & 2.33 & 60.44 & 40.74 & 101.18 & 14.33 & 1.06 & 15.39 & 46.59 & 61.98  \\
VC   & 10.62 & 0.93 & 11.55 & 20.65 & 32.20 &  3.29  & 0.50 & 3.79 & 22.56 & 26.35  \\
NbC  & 18.95 & 0.73 &  19.68 & 30.23 & 49.91 & 5.79 & 0.33 & 6.12 & 32.70 & 38.82  \\
TaC  & 26.46 & 0.88 & 27.34 & 38.17 & 65.51 & 7.49 & 0.40 & 7.89 & 41.91 & 49.80  \\
SiC  & 415.33 & 1.34 & 416.67 & - & 416.67  & 89.69 & 1.36 & 91.05 & - & 91.05 \\
\hline
\end{tabular}
}
\label{tab1}
\end{table*}

For the Group~IV transition-metal carbides (TiC, ZrC, and HfC), the total thermal conductivity is obtained by summing the lattice and electronic contributions:
\begin{equation}
\kappa_{\mathrm{total}}=
\kappa_{\mathrm{ph}}+\kappa_{\mathrm{el}},
\end{equation}
where $\kappa_{\mathrm{ph}}$ and $\kappa_{\mathrm{el}}$ denote the lattice and electronic thermal conductivities, respectively. As shown in Figure~\ref{fig5}(a--c), the lattice contribution accounts for approximately 60\% of the total thermal conductivity near room temperature. With increasing temperature, enhanced anharmonic phonon--phonon scattering reduces $\kappa_{\mathrm{ph}}$, which exhibits an approximately inverse-temperature dependence over the investigated range. Consequently, the lattice contribution decreases to about 30\% at $1000~\mathrm{K}$ and approximately 10\% at $3500~\mathrm{K}$.

In contrast, the relative contribution from electronic heat transport increases with temperature. The lattice and electronic contributions become comparable at approximately $600~\mathrm{K}$ for the investigated Group~IV carbides, although the precise crossover temperature depends on the compound. The increasing relative importance of electronic transport partly compensates for the decreasing lattice contribution, resulting in a weaker temperature dependence of the total thermal conductivity than that of either contribution considered separately.

The Group~V transition-metal carbides (VC, NbC, and TaC) exhibit temperature-dependent trends in thermal conductivity similar to those of the Group~IV carbides, but their calculated lattice thermal conductivities are substantially lower (Figure~\ref{fig5}(d--f) and Table~\ref{tab1}). The lattice contribution shows an approximately inverse-temperature dependence, consistent with the enhanced phonon--phonon scattering at elevated temperatures. At room temperature, the lattice contribution accounts for approximately 40\% of the total thermal conductivity, compared with approximately 60\% for the Group~IV carbides. This fraction decreases further as temperature increases, reaching around 10\% at $3500~\mathrm{K}$.

The lower lattice thermal conductivity of the Group~V carbides coincides with the pronounced Kohn anomalies identified in their calculated phonon dispersions. These anomalies modify the phonon frequencies in selected regions of the Brillouin zone and may influence lattice heat transport through changes in phonon group velocities. The differences in the phonon dispersions are therefore a possible contributor to the lower lattice thermal conductivity calculated for the Group~V carbides. 
Electronic heat transport dominates the total thermal conductivity of the Group~V carbides over the investigated temperature range. As in the Group~IV carbides, the increasing relative contribution from electronic transport partly compensates for the decreasing lattice contribution, resulting in a comparatively weaker temperature dependence of the total thermal conductivity at elevated temperatures.


\begin{table*}
\centering
\caption{Fitting parameters for phonon, electronic, and total thermal conductivity of MC (M=Ti, Zr, Hf, V, Nb, and Ta) and SiC.} 
\resizebox{1.0\columnwidth}{!}{%
\begin{tabular}{ll S S S S S S S}
\toprule
Contribution & Parameter &  {TiC}  & {ZrC} & {HfC} & {VC} & {NbC} & {TaC} & {SiC} \\
\midrule
\multirow{3}{*}{Phonon ($\kappa_{ph}$)} 
& $A (\times 10^{4})$ & 4.416   & 2.211 & 6.638 & 0.174 & 0.447 & 1.132 & 97.767 \\
& $n$ & 1.130  & 1.064 & 1.234 & 0.874 & 0.949 & 1.058 & 1.368 \\
& $B$ & 1.544  & 0.659 & 1.888 & -0.347 & -0.211 & 0.296 & 10.000 \\

\addlinespace

\multirow{3}{*}{Electronic ($\kappa_{el}$)} 
& $a$ & 0.006  & 0.007 & 0.008 & 0.001 & 0.0003 & 0.001    \\
& $b$ & 37.107  & 35.332 & 38.755 & 22.086 & 33.260 & 42.291   \\
& $c (\times 10^{3})$ & -0.453  & -1.536 & -0.498 & -0.629 & -0.949 & -1.292  \\


\addlinespace


\bottomrule
\end{tabular}
}
\label{tab3}
\end{table*}

Overall, the calculated results indicate that thermal transport in the investigated transition-metal carbides reflects the combined contributions of phonons and conduction electrons. At relatively low temperatures, acoustic phonons provide a substantial contribution to lattice heat transport, whereas increasing anharmonic phonon--phonon scattering suppresses the lattice thermal conductivity as temperature rises. The electronic contribution becomes increasingly important in relative terms, particularly at elevated temperatures. To describe the calculated temperature dependence of the lattice and electronic contributions over the investigated range, we employ the following empirical relations:
\begin{equation}
\kappa_{\mathrm{ph}}(T)=\frac{A}{T^n}+B,
\label{eq:phfit}
\end{equation}
and
\begin{equation}
\kappa_{\mathrm{el}}(T)=aT+b+\frac{c}{T}.
\label{eq:elfit}
\end{equation}
In the first expression, $A$ is a scaling parameter, $n$ characterizes the temperature dependence, and $B$ represents a temperature-independent offset. In the second expression, $a$ and $b$ describe the linear and constant contributions, respectively, while the $c/T$ term accounts phenomenologically for additional temperature dependence. The fitted curves reproduce the calculated thermal conductivities with coefficients of determination of approximately $R^2=0.99$. The fitting parameters are summarized in Table~\ref{tab3}, and the corresponding fitted curves are shown in Figure~\ref{fig5}. These relations provide compact descriptions of the calculated trends over the temperature range considered; they are not intended as universal transport laws.

\begin{figure}
\centering{\includegraphics[width=0.6\linewidth]{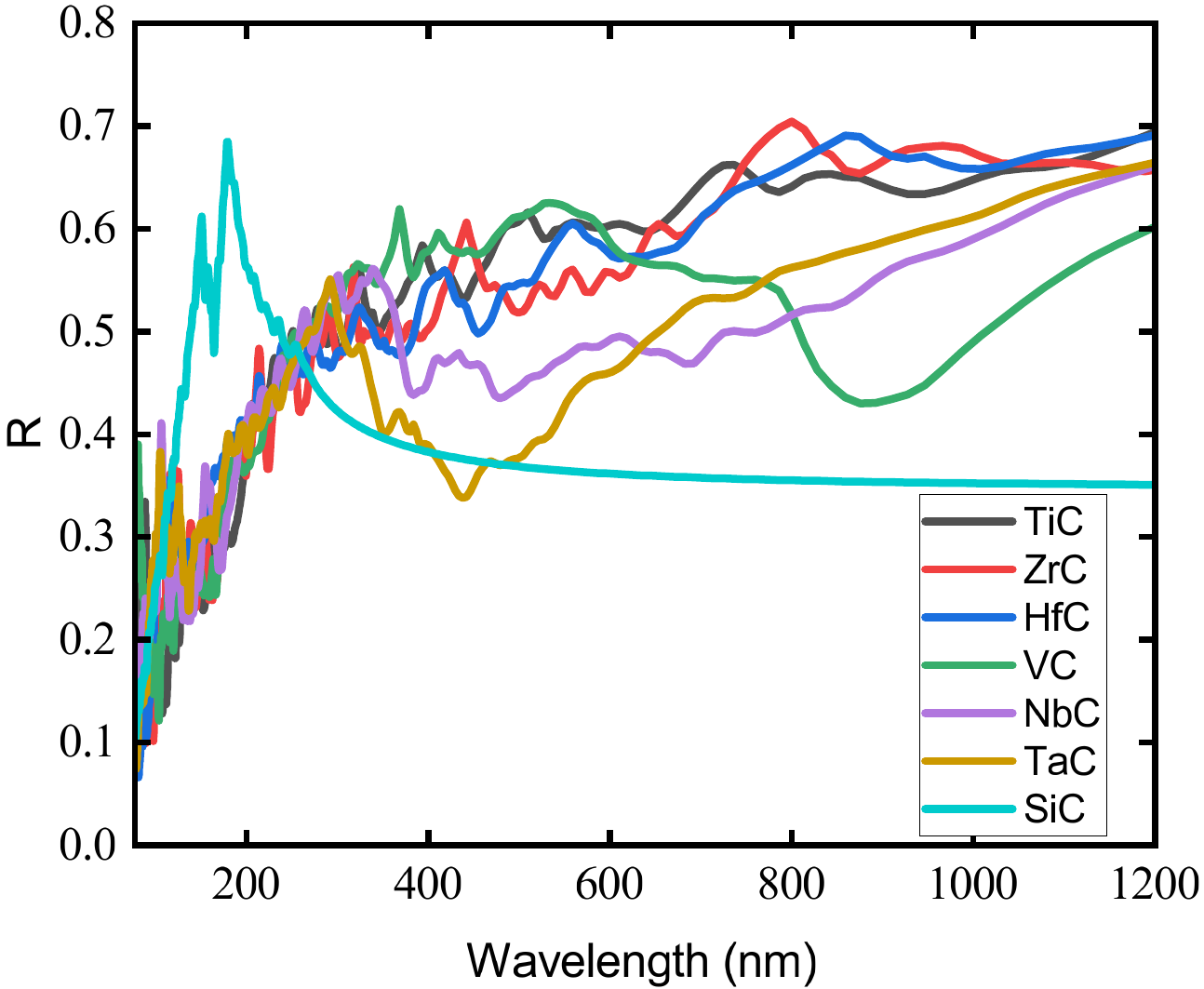}}
\caption{Calculated reflectivity spectra of TiC, ZrC, HfC, VC, NbC, TaC, and SiC as functions of wavelength. In the
1000--1200~nm interval, the transition-metal carbides exhibit reflectivities of approximately 0.55--0.69, whereas SiC
exhibits a lower reflectivity of approximately 0.35.}
\label{fig6}
\end{figure}

\subsection{Optical Properties}
The optical response of a material to incident electromagnetic radiation can be characterized by the absorption coefficient $\alpha(\omega)$, refractive index $n(\omega)$, extinction coefficient $k(\omega)$, and reflectivity $R(\omega)$. These quantities can be obtained from the real and imaginary parts of the complex dielectric function, $\varepsilon(\omega)=\varepsilon_r(\omega)+i\varepsilon_i(\omega)$. The absorption coefficient, refractive index, extinction
coefficient, and reflectivity are given by~\cite{yadav2025high} 
\begin{equation}
\alpha(\omega)=\frac{\sqrt{2}\,\omega}{c}
\left[
\sqrt{\varepsilon_r^2(\omega)+\varepsilon_i^2(\omega)}
-\varepsilon_r(\omega)
\right]^{1/2},
\end{equation}
\begin{equation}
n(\omega)=
\left[
\frac{\sqrt{\varepsilon_r^2(\omega)+\varepsilon_i^2(\omega)}
+\varepsilon_r(\omega)}{2}
\right]^{1/2},
\end{equation}
\begin{equation}
k(\omega)=
\left[
\frac{\sqrt{\varepsilon_r^2(\omega)+\varepsilon_i^2(\omega)}
-\varepsilon_r(\omega)}{2}
\right]^{1/2},
\end{equation}
and
\begin{equation}
R(\omega)=
\frac{[n(\omega)-1]^2+k^2(\omega)}
{[n(\omega)+1]^2+k^2(\omega)}.
\end{equation}
Here, $\omega$ is the angular frequency of the incident radiation and $c$ is the speed of light in vacuum. The reflectivity expression assumes normal incidence at a planar interface between vacuum and a nonmagnetic material. Angular frequency and wavelength are related by $\omega=2\pi c/\lambda$, where $\lambda$ denotes the wavelength. The calculated real and imaginary components of the dielectric function for the Group~IV transition-metal carbides, SiC, and the Group~V transition-metal carbides are presented in Figures~S5 and~S6 of the Supplementary Material.

The calculated reflectivity spectra of TiC, ZrC, HfC, VC, NbC, TaC, and SiC are shown in Figure~\ref{fig6} as functions of wavelength. In the wavelength range of 1000--1200 nm, the transition-metal carbides exhibit reflectivities of approximately 0.55--0.69, consistent with their metallic electronic character [Table S1, Supplementary Material]. Among these materials, TaC, HfC, TiC, and ZrC exhibit reflectivities of approximately 0.69, 0.68, 0.67, and 0.67, respectively, whereas the corresponding values for NbC and VC are approximately 0.63 and 0.55. In contrast, SiC exhibits a lower reflectivity of approximately 0.35 in the same wavelength interval, consistent with its semiconducting electronic structure. These results indicate that the optical response varies among the investigated materials, even within the same wavelength range. The corresponding optical absorption coefficient and refractive-index spectra are presented in Figures~S7(a) and S7(b), respectively, in the Supplementary Material.

The reflectivity is also an important input to the subsequent laser-heating calculations, in which the absorbed surface power flux is approximated as $q_{\mathrm{abs}}=(1-R)I$, where $I$ is the incident laser intensity. Accordingly, differences in reflectivity affect the fraction of incident radiation that enters the material and must be considered alongside thermal transport properties when comparing predicted temperature responses. Reflectivity alone, however, does not determine the thermal response or establish the resistance of a material to laser-induced damage.

\subsection{Thermal Response of Carbides under Laser Irradiation}
The thermal and optical properties calculated from first principles are used to examine the temperature response of the investigated carbides under laser irradiation. In particular, the electronic contribution to thermal conductivity is important for describing heat transport in the metallic transition-metal carbides and should be considered alongside the lattice contribution. To compare the thermal responses under different geometric assumptions, we employ two continuum models: a one-dimensional (1D) model representing a semi-infinite solid and a three-dimensional (3D) model representing a finite plate. The 1D model provides an analytical estimate of the surface-temperature evolution, whereas the 3D model accounts for finite specimen dimensions and convective heat loss. Together, these models provide a comparative assessment of laser-induced heating under the specified irradiation conditions.  

The detailed analytical solution for the 1D model is provided in Section S2 of the Supplementary Material; we present the key results here. For the semi-infinite solid, the surface temperature at time $t$ is expressed as:
\begin{equation}
    T(0,t) = T_0 + \frac{2q}{\kappa}\sqrt{\frac{\alpha t}{\pi}}~,
\label{1deq1}
\end{equation}
where $T_0$ is the initial temperature (taken as the room temperature, 300 K) and $\kappa$ is the thermal conductivity (electron and phonon parts together). The absorbed surface heat flux is approximated as, 
\begin{equation}
q=(1-R)I,    
\end{equation}
where $R$ is the reflectivity and $I$ is the incident laser intensity: higher reflectivity reduces the laser absorption. Thermal diffusivity is calculated using
\begin{equation}
\alpha=\frac{\kappa}{\rho C_{p}},
\label{eq:alpha}
\end{equation}
where $\rho$ is the density and $C_{p}$ is the average heat capacity at constant pressure. 

Since the material properties depend on temperature, to evaluate Eq.~\ref{1deq1} and Eq.~\ref{eq:alpha}, representative values of thermal conductivity and heat capacity are obtained by averaging their temperature-dependent values between the initial temperature $T_0$ and the melting temperature $T_m$,
\begin{equation}
\kappa_{av} = \frac{1}{(T_m-T_0)}\int_{T_0}^{T_m} \kappa(T) \,dT,
\label{eq:1d13}
\end{equation}
\begin{equation}
C_{p,av} = \frac{1}{(T_m-T_0)}\int_{T_0}^{T_m} C_p(T) \,dT.
\label{eq:1d14}
\end{equation}
Thermal conductivity values are taken from our first-principles calculations [Eq.~\ref{eq:phfit} and Eq.~\ref{eq:elfit}]. The Shomate equation is used to estimate the heat capacity at constant pressure,
\begin{equation}
C_p = A + Bt^{'} + Ct^{'2} + Dt^{'3} + E/t^{'2},
\label{eq:1d15}
\end{equation}
where $t'$=T/1000 (T is temperature in K). Values of the constants ($A, B, C, D, E$) are provided in Table S2, Supplementary Material.

\begin{table*}
\centering
\caption{The density ($\rho$, kg m$^{-3}$), melting point (T$_m$, K), heat capacity at constant pressure (C$_{p,av}$, J kg$^{-1}$ K$^{-1}$), thermal conductivity ($\kappa_{av}$, W m$^{-1}$ K$^{-1}$), and reflectivity ($R$) of group transition metal carbides, SiC, Al, and Cu. Also shown is the surface temperature (K) after 5 s of exposure to laser radiation of 1 kW cm$^{-2}$ intensity, calculated using the 1D ($T_{t=5}^{1D}$) model, assuming a semi-infinite thickness. The 3D model assumes a beam radius of 0.8 mm and two different plate thicknesses: 0.3 mm ($T_{t=5}^{3D_1}$) and 3 mm ($T_{t=5}^{3D_2}$), both having the same cross-sectional area of 4 mm$^2$.}
\begin{tabular}{ c c c c c c c c c}
\hline
Material & $\rho$ & T$_m$ & C$_{p,av}$ & $\kappa_{av}$ & R & $T_{t=5}^{1D}$ & $T_{t=5}^{3D_1}$ & $T_{t=5}^{3D_2}$ \\
\hline
TiC  & 4942 & 3523 & 963 & 64 & 0.67 & 777 & 1601 & 481   \\
ZrC  & 6678 & 3803 & 550 & 60 & 0.67 & 861 & 1903 & 523 \\
HfC  & 12708 & 4201 & 280 & 65 & 0.68 & 831 & 1893 & 519 \\
VC  & 5988 & 2921 & 1092 & 26 & 0.55 & 1171 & 1797 & 578 \\
NbC & 8000 & 3773 & 521 & 38 & 0.63 & 1042 & 1976 & 556 \\
TaC & 14944 & 4153 & 315 & 49 & 0.69 & 815 & 1554 & 483\\
SiC & 3216 & 2873 & 1245 & 82 & 0.35 & 1205 & 3225 & 690 \\
Al  & 2700 & 933 & 900 & 235 & 0.76 & 553 & 1847 & 501 \\
Cu & 8980 & 1350 & 385 & 420 & 0.8 & 432 & 1281 & 419 \\
\hline
\end{tabular}
\label{tabS3}
\end{table*}

The resulting average properties, together with the density, melting temperature, and reflectivity, are
summarized in Table~\ref{tabS3}. Note that, other than melting point and heat capacity, the rest of the parameters are taken from our \textit{ab initio} calculations. Also shown are the properties of Al and Cu taken from the literature~\cite{tsumura2025threshold}. The table shows the surface temperatures calculated using the 1D model, which assumes a semi-infinite thickness. The results of Al and Cu are included for comparison with the transition-metal carbides and SiC. Under the specified irradiation conditions of $1~\mathrm{kW\,cm^{-2}}$ for 5~s, the 1D model predicts surface temperatures ($T_{t=5}^{1D}$) below the corresponding melting temperatures for all the materials considered [see Figure S8, Supplementary Materials for further details]. These results provide a comparison of their thermal responses within the assumptions of the semi-infinite conduction model.

The detailed analytical solution for the 3D model is provided in Section S3 of the Supplementary Material; we present the key results here. In this model, the temperature field is calculated for a finite plate, accounting for heat conduction along all three spatial directions and convective heat loss at the specified boundaries. The temperature distribution in 3D at any location $(x,y,z)$ and time $t$ is given by:
\begin{equation}
\begin{split}
T(x,y,z,t)=T_0+ \\
\frac{I_0(1-R)\pi w^2}{kL_xL_yL_z}\sum_{n=1}^{\infty}\sum_{p=1}^{\infty}\sum_{m=1}^{\infty}\frac{\eta_n \mathrm{exp}\left[-\frac{w^2}{8}(\lambda_m^2+\beta_p^2)\right]}{(\frac{2H}{L_x}+H^2+\lambda_m^2)(\frac{2H}{L_y}+H^2+\beta_p^2)(\frac{2H}{L_z}+H^2+\eta_n^2)}\\
\left[\lambda_m\cos(\lambda_m x)+H\sin(\lambda_m x)\right]\left[\beta_p\cos(\beta_p y)+H\sin(\beta_p y)\right] 
    \left[\eta_n \cos(\eta_n z)+H \sin(\eta_n z)\right]\\
    \left[\lambda_m \mathrm{Re}[P] + H\mathrm{Im}[P]\right]\left[\beta_p \mathrm{Re}[Q] + H\mathrm{Im}[Q]\right]\\
    \left[\frac{1-\mathrm{exp}\left\{-\alpha(\lambda_m^2 + \beta_p^2 + \eta_n^2)t\right\}}{\lambda_m^2 + \beta_p^2 + \eta_n^2}\right],
\end{split}
\label{eq:3df46}
\end{equation}

\begin{equation}
    P = \mathrm{exp}\left\{\frac{iL_x\lambda_m}{2}\right\}\left[\mathrm{erf}\left\{\frac{L_x}{\sqrt{2}w}-i\frac{\lambda_m w}{2\sqrt2}\right\}+\mathrm{erf}\left\{\frac{L_x}{\sqrt{2}w}+i\frac{\lambda_m w}{2\sqrt2}\right\}\right],
    \label{eq:3df43}
\end{equation}
\begin{equation}
    Q = \mathrm{exp}\left\{\frac{iL_y\beta_p}{2}\right\}\left[\mathrm{erf}\left\{\frac{L_y}{\sqrt{2}w}-i\frac{\beta_p w}{2\sqrt2}\right\}+\mathrm{erf}\left\{\frac{L_y}{\sqrt{2}w}+i\frac{\beta_p w}{2\sqrt2}\right\}\right].
    \label{eq:3df44}
\end{equation}
Here $I_0$ is the average intensity, $R$ is reflectivity, $w$ is beam radius, $L_x$,$L_y$, and $L_z$ are the sizes of the 3D geometry, $\lambda$, $\beta$, and $\eta$ are the eigenvalues, and $H=h/\kappa$, where $h$ = convective heat transfer coefficient from the solid to the air. Two different material thicknesses are chosen: a thin plate of 0.3 mm and a thick plate of 3 mm, both with a cross-sectional area of 4 mm$^2$. The calculations are performed for laser irradiation at a wavelength of 1000~nm, an incident intensity of $1~\mathrm{kW\,cm^{-2}}$, and a beam radius of 0.8~mm. A convective heat-transfer coefficient of $40~\mathrm{W\,m^{-2}\,K^{-1}}$ is used. These conditions are held fixed to compare the temperature responses of the materials under investigation and to examine the influence of specimen thickness within the adopted continuum model. 

\begin{figure}[h]
\centering{\includegraphics[width=1\linewidth]{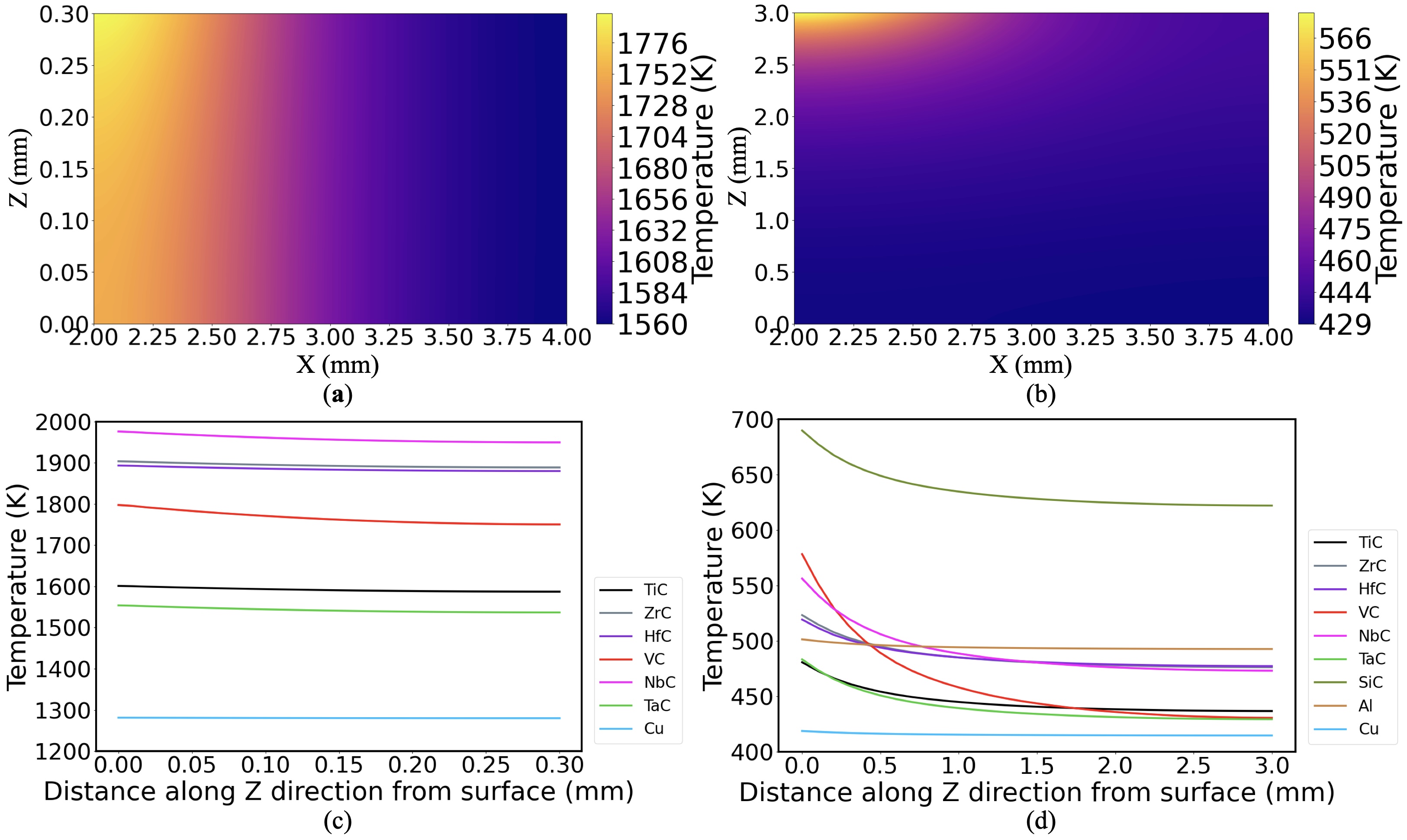}}
\caption{(a) XZ section of the temperature profile in VC calculated for 5 s exposure to laser radiation of 1000 nm wavelength, 1 kW cm$^{-2}$ intensity, 0.8 mm beam radius, $h=40$ Wm$^{-2}$K$^{-1}$; calculated for a 3D plate with a cross-sectional area of 4 mm$^2$ and thickness of (a) 0.3 mm and (b) 3 mm. Temperature versus distance along the z-direction for a plate thickness of (c) 0.3 mm and (d) 3 mm. Also see Figure S9 and S10, Supplementary Materials.}
\label{fig9}
\end{figure}

Figure~\ref{fig9} shows the temperature distribution in the XZ plane for VC and the temperature profiles along the
thickness direction for the investigated carbides MC (M = Ti, Zr, Hf, V, Nb, Ta, Si) and metals (Al, Cu). The predicted temperatures for the 0.3~mm plates of SiC and Al exceed their respective melting temperatures. These cases are therefore not included in the corresponding solid-state temperature maps. Because the present model does not explicitly describe melting or the associated latent heat, the calculated temperatures above melting should be interpreted as indicators of the limits of the solid-state model rather than as quantitative predictions of the temperature during or after melting. 

The differences between the 1D and 3D results in Table~\ref{tabS3} reflect the different geometric representations and heat-transfer conditions adopted by the two models: heat can diffuse in 3 directions in 3D, whereas in 1D it can diffuse only in 1 direction. The temperature predicted by the 3D model increases significantly as the system thickness decreases, because a material of much smaller size receives the same amount of heat flux. For example, the predicted surface temperature of TiC decreases from 1601~K for the 0.3~mm plate to 481~K for the 3~mm plate. For SiC, the corresponding values are 3225 K and 690 K. These results demonstrate the substantial influence of finite specimen geometry on the calculated thermal response under the specified irradiation conditions. These results also highlight the importance of the 3D model, as a 1D model with semi-infinite geometry cannot accurately capture the effect of system size.

The comparison in Table~\ref{tabS3} shows that Cu exhibits the lowest predicted surface temperature among the investigated materials for both plate thicknesses. However, the absolute surface temperature alone does not capture the influence of melting temperature on the thermal response of the carbides. For the 0.3~mm plates, the predicted surface temperatures of the transition-metal carbides remain substantially below their respective melting temperatures, whereas the predicted temperature of Cu approaches its melting point. For the 3~mm plates, the predicted surface temperatures remain below the corresponding melting temperatures for all materials considered.

\begin{figure*}
\centering{\includegraphics[width=1\linewidth]{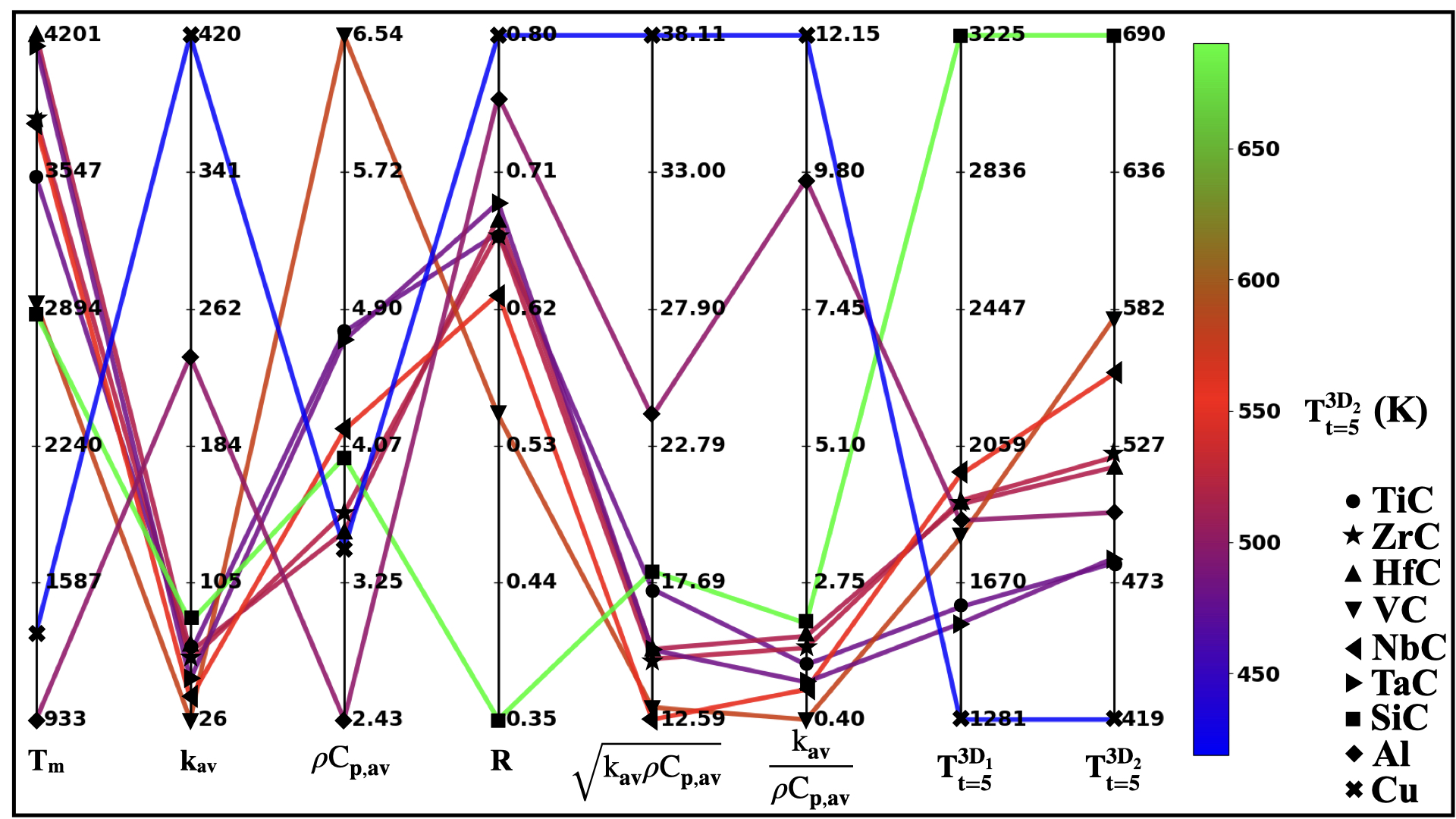}}
\caption{Melting point (T$_m$, K), thermal conductivity ($\kappa_{av}$,  W m$^{-1}$ K$^{-1}$), product of density and heat capacity at constant pressure ($\rho C_{p,av}$ $\times 10^6$ J m$^{-3}$ K$^{-1}$), reflectivity (R), thermal effusivity ($\sqrt{\kappa_{av}\rho C_{p,av}}$, $\times 10^3$ W s$^{0.5}$m$^{-2}$K$^{-1}$), thermal diffusivity ($\alpha=\frac{\kappa_{av}}{\rho C_{p,av}}$, $\times 10^{-5}$ m$^2$ s$^{-1}$), surface temperature after 5 s exposure to LASER radiation of 1 kW cm$^{-2}$ intensity with 1000 nm wavelength and beam radius as 0.8 mm calculated using 3D models for plate with 0.3 mm ($T_{t=5}^{3D_1}$, K) and 3 mm thickness ($T_{t=5}^{3D_2}$, K).}
    \label{fig8}
\end{figure*}

Figure~\ref{fig8} clearly illustrates the difference between the two classes, metals like Al, Cu, and the carbides. Specifically, two properties, thermal diffusivity $\left(\alpha=\frac{\kappa_{av}}{\rho C_{p,av}}\right)$ and thermal effusivity $\left(\sqrt{\kappa_{av}\rho C_{p,av}}\right)$, are particularly important for thermal transport. Thermal diffusivity characterizes the rate at which temperature disturbances spread through a material, whereas thermal effusivity describes its capacity to exchange heat with its surroundings. Higher values of thermal effusivity and diffusivity lead to a lower rise in surface temperature. Cu has significantly higher thermal effusivity and diffusivity values than carbides, and even the other metal, Al, which has an intermediate value because of its lowest $\rho C_{p,av}$ value among all the materials studied here. Because of its highest effusivity and diffusivity, the lowest surface temperature rise is observed in Cu for both low- and high-thickness plates. 

The other metal, Al, performs well at a plate thickness of 3 mm, but the surface temperature of the 0.3 mm thickness exceeds its melting point, as Al has a low melting point. Among the carbides, only SiC behaves differently, with surface temperature rising rapidly compared to other carbides, which is mainly a consequence of the low reflectivity of SiC. Although Cu seems to be outperforming carbides in terms of surface temperature rise, one should note that the surface temperature is $\sim 0.95T_m$ of Cu in a 0.3 mm thick plate. Although surface temperatures of 0.3 mm thick carbide plates are higher than those of the Cu plate, they are still $\sim 0.6 T_m$  or even less. Thus, the results suggest that the high melting temperatures of the transition-metal carbides may provide an advantage when evaluating their proximity to melting under the specified irradiation conditions. Overall, Figure~\ref{fig8} indicates that material selection under laser irradiation cannot be based on thermal conductivity or reflectivity alone; the relevant properties must be considered
together with the geometry.

\section{Conclusion}
In this work, we investigated the electronic structure, lattice dynamics, thermal transport, and optical properties of Group IV transition-metal carbides (TiC, ZrC, and HfC), Group V transition-metal carbides (VC, NbC, and TaC), and SiC using first-principles calculations. The calculated electronic structures indicate metallic behavior in the transition-metal carbides, whereas SiC exhibits semiconducting behavior with an indirect band gap. The thermal-transport calculations reveal distinct trends in the two carbide groups: the Group IV carbides exhibit a crossover from predominantly lattice-mediated heat transport at lower temperatures to an increasingly important electronic contribution at elevated temperatures, whereas electronic heat transport dominates more strongly in the Group V carbides over the investigated temperature range. The Group V carbides also exhibit substantially lower calculated lattice thermal conductivity than the Group IV carbides. In contrast, SiC exhibits the highest lattice thermal conductivity among the investigated materials, consistent with the contribution of its relatively high phonon group velocities and phonon lifetimes.

Fermi-surface analysis indicates relatively weak nesting in the Group IV carbides, whereas more pronounced nesting features and Kohn anomalies are observed in the Group V carbides. These differences are associated with modifications to the phonon dispersion and contribute to the contrasting behavior of lattice thermal transport. The optical calculations further show that the transition-metal carbides exhibit relatively high reflectivity, approximately 0.55–0.69 in the wavelength interval of 1000–1200 nm, compared with approximately 0.35 for SiC. These optical differences affect the fraction of incident laser radiation that enters the material under the adopted surface-flux approximation. Together, the results highlight the importance of considering electronic structure, lattice dynamics, thermal transport, and optical response when comparing the thermal behavior of these materials.

The continuum laser-heating calculations demonstrate that the predicted temperature response depends on both material properties and specimen geometry. Under an incident intensity of $1~\mathrm{kW\,cm^{-2}}$ for 5~s, the 1D semi-infinite model predicts surface temperatures below the corresponding melting temperatures for all investigated materials. In the 3D model, reducing the plate thickness from 3~mm to 0.3~mm substantially increases the predicted surface temperature. The calculated temperatures for the thin SiC and Al plates exceed their respective melting temperatures, whereas the temperatures predicted for the 3~mm plates remain below the corresponding melting points. Although Cu exhibits the lowest absolute surface temperature in both plate geometries, the high melting temperatures of the transition-metal carbides provide a different basis for comparing their thermal response. These results emphasize the combined roles of optical reflectivity, thermal transport properties, melting temperature, and specimen geometry in evaluating candidate materials for laser-exposed applications.

\begin{section}{Acknowledgments}
The authors gratefully thank the Defense Laboratory (DRDO), Jodhpur, for research funding (DGNSM/06/6031/DLJ-351/CARS/IIT-Kanpur). This research used computational resources at PARAM Rudra 20 PF HPC System provided by the Centre for Development of Advanced Computing (C-DAC) under the National Supercomputing Mission (NSM), India's flagship initiative jointly steered by the Ministry of Electronics and Information Technology (MeitY) and the Department of Science and Technology (DST).
\end{section}

\appendix 
\section{Lattice Thermal Conductivity}
\label{app:ltc}
The lattice thermal conductivity is calculated using PHONO3PY~\cite{togo2008first, PhysRevB.84.094302, PhysRevB.91.094306},
which employs the finite-displacement method to obtain the interatomic force constants (IFCs) and subsequently solves the
phonon Boltzmann transport equation (BTE). The second-order and third-order IFCs are calculated using $2\times2\times2$ supercells containing 64 atoms. The corresponding first-principles force calculations are performed using an $8\times8\times8$ $\mathbf{q}$-point grid. The second-order IFCs are used to determine the phonon frequencies, eigenvectors, and group velocities, whereas the third-order IFCs are employed to calculate the anharmonic three-phonon scattering rates. The resulting phonon properties, including the mode-resolved heat capacities, group velocities, and phonon relaxation times, are used to evaluate the lattice thermal conductivity. The calculations explicitly account for three-phonon scattering, including both normal and Umklapp processes.

Within the single-mode relaxation-time approximation (RTA), the lattice thermal-conductivity tensor is expressed as
\begin{equation}
\kappa_{\mathrm{ph}}^{ab}
=
\frac{1}{V N_{\mathbf q}}
\sum_{\lambda}
C_{\lambda}
v_{\lambda}^{a}
v_{\lambda}^{b}
\tau_{\lambda},
\label{eqn:ph-ph}
\end{equation}
where $\lambda=(\mathbf q,\nu)$ denotes a phonon mode characterized by the wave vector $\mathbf q$ and branch index $\nu$, $V$ is the primitive-cell volume, and $N_{\mathbf q}$ is the number of sampled wave vectors in the Brillouin zone. Here, $C_{\lambda}$ is the mode-resolved heat capacity, $v_{\lambda}^{a}$ and $v_{\lambda}^{b}$ are the components of the phonon group velocity along the Cartesian directions $a$ and $b$, respectively, and $\tau_{\lambda}$ is the phonon relaxation time associated with anharmonic scattering. The relaxation times are determined from the three-phonon scattering rates calculated using the third-order IFCs. Normal processes conserve phonon crystal momentum, whereas Umklapp processes transfer crystal momentum to the reciprocal lattice, thereby contributing to the resistance to heat transport~\cite{broido2007intrinsic}.

\section{Electronic Thermal Conductivity}
\label{app:etc}
The electronic thermal conductivity $\kappa_{\mathrm{el}}$ is calculated using the BTE and Onsager relations as~\cite{giustino2017rmp}

\begin{equation}
[\sigma]_{ab} = -\frac{e^2 n_s}{V} \sum_{i \mathbf{k}} \frac{d f_{i \mathbf{k}}}{d \varepsilon} \, v_{i \mathbf{k}}^{a} v_{i \mathbf{k}}^{b} \tau_{i \mathbf{k}}^{el},
\label{eqn:sigma}
\end{equation}
\begin{equation}
[\sigma S]_{ab} = -\frac{e n_s}{V T} \sum_{i \mathbf{k}} (\varepsilon_{i \mathbf{k}} - \mu) \frac{d f_{i \mathbf{k}}}{d \varepsilon} \, v_{i \mathbf{k}}^{a} v_{i \mathbf{k}}^{b} \tau_{i \mathbf{k}}^{el},
\label{eqn:sigmaS}
\end{equation}
\begin{equation}
[K]_{ab} = -\frac{n_s}{V T} \sum_{i \mathbf{k}} (\varepsilon_{i \mathbf{k}} - \mu)^2 \frac{d f_{i \mathbf{k}}}{d \varepsilon} \, v_{i \mathbf{k}}^{a} v_{i \mathbf{k}}^{b} \tau_{i \mathbf{k}}^{el}.
\label{eqn:K}
\end{equation}
Here $\mathbf{k}$ is the electronic wave vector corresponding to the band index $i$, $n_s$ is the spin degeneracy, $V$ is the volume of the unit cell, and $e$ is the elementary charge. The term $f_{i\mathbf{k}}$ is the equilibrium Fermi--Dirac distribution function, while $\varepsilon_{i\mathbf{k}}$ and $\mu$ are the electron energy and the chemical potential, respectively. The quantity $v_{i\mathbf{k}}^{a}$ is the $a$ component of the electron group velocity and $\tau_{i\mathbf{k}}^{el}$ denotes the electron relaxation time. The indices $a$ and $b$ indicate the Cartesian components of the transport tensors~\cite{tong2019comprehensive,ponce2020rpp}. The Seebeck coefficient follows from Eqs.~(\ref{eqn:sigma}) and (\ref{eqn:sigmaS}) as $S = [\sigma]^{-1}[\sigma S]$, and the tensor $K$ then gives the electronic
thermal conductivity as
\begin{equation}
\kappa_{\mathrm{el}} = K - S \sigma S T,
\label{eqn:kappael}
\end{equation}
where $T$ is the temperature. For the cubic carbides considered here, all three transport tensors are isotropic and are reduced to scalars by taking one third of their trace.

Equations~(\ref{eqn:sigma})--(\ref{eqn:K}) are written in the self-energy relaxation-time approximation (SERTA)~\cite{giustino2017rmp,ponce2020rpp}, in which the deviation of the occupation from equilibrium is taken to be $F_{i\mathbf{k}}^{a} = -\left(df_{i\mathbf{k}}/d\varepsilon\right) v_{i\mathbf{k}}^{a}\,\tau_{i\mathbf{k}}^{el}$, that is, strictly proportional to the group velocity. In this work, the linearized BTE is instead solved iteratively (IBTE)~\cite{li2015prb,ponce2018prb}, for which
\begin{equation}
F_{i \mathbf{k}}^{a} = \tau_{i \mathbf{k}}^{el}
\left[ -\frac{d f_{i \mathbf{k}}}{d \varepsilon}\, v_{i \mathbf{k}}^{a}
\;+\; \sum_{j \mathbf{k}'} W_{i \mathbf{k} \rightarrow j \mathbf{k}'}\,
F_{j \mathbf{k}'}^{a} \right],
\label{eqn:ibte}
\end{equation}
where $W_{i\mathbf{k} \rightarrow j\mathbf{k}'}$ is the electron--phonon
transition rate between states connected by absorption or emission of a phonon, and $\sum_{j\mathbf{k}'} W_{i\mathbf{k} \rightarrow j\mathbf{k}'} = 1/\tau_{i\mathbf{k}}^{el}$. The second term is the scattering-in contribution: an electron scattered out of $i\mathbf{k}$ arrives in a state that is itself displaced from equilibrium, so part of the current is transferred rather than destroyed. Discarding it recovers the SERTA expressions above. Because the sum collects $F_{j\mathbf{k}'}$ from states whose velocities point in different directions, $F_{i\mathbf{k}}$ is in general not parallel to $v_{i\mathbf{k}}$ and therefore cannot be represented by any scalar relaxation time~\cite{claes2025nrp}. The transport tensors follow from Eqs.~(\ref{eqn:sigma})--(\ref{eqn:K}) with $-\left(df_{i\mathbf{k}}/d\varepsilon\right) v_{i\mathbf{k}}^{a} \tau_{i\mathbf{k}}^{el}$ replaced by $F_{i\mathbf{k}}^{a}$. Equation (\ref{eqn:ibte}) is solved by direct iteration starting from the SERTA solution~\cite{ponce2016cpc,lee2023npj}. 

All quantities entering Eqs.~(\ref{eqn:sigma})--(\ref{eqn:ibte}) are obtained from first principles, with no adjustable parameters: in particular, $\tau_{i\mathbf{k}}^{el}$ is not fitted but computed from the lowest-order electron--phonon self-energy~\cite{giustino2017rmp}, its inverse being the Fermi golden-rule sum of the transition rates $W_{i\mathbf{k}\rightarrow j\mathbf{k}'}$ over all final states, phonon branches and both absorption and emission channels. The calculations use \textsc{Quantum ESPRESSO}~7.0~\cite{giannozzi2009jpcm,giannozzi2017jpcm} for the ground state and lattice dynamics and \textsc{EPW}~5.4.1~\cite{ponce2016cpc,lee2023npj} for the Wannier interpolation and transport.

The carbides are treated in the two-atom rocksalt primitive cell at the experimental lattice constant. 
Metallic occupations are described by Fermi--Dirac smearing of width $0.01$~Ry, and the charge density is converged to $10^{-12}$~Ry on an $8\times8\times8$ Monkhorst--Pack $\mathbf{k}$ grid~\cite{monkhorst1976prb}.
Dynamical matrices and the self-consistent variation of the Kohn--Sham potential are obtained by density-functional perturbation theory~\cite{baroni2001rmp} on a commensurate $8\times8\times8$ $\mathbf{q}$ grid ($29$ irreducible points, convergence threshold $10^{-14}$) and Fourier-transformed to real-space interatomic force constants, to which a crystal acoustic sum rule is applied. A non-self-consistent calculation with $28$ bands on the full $512$-point $8\times8\times8$ $\mathbf{k}$ grid supplies the wavefunctions for the Wannier transformation~\cite{pizzi2020jpcm}. Nine maximally localized Wannier functions~\cite{marzari2012rmp} are constructed per compound from $M$:$d$ and C:$sp^{3}$ projections in $500$ minimization steps, excluding the metal semi-core bands while retaining the C~$2s$ band [Section~S4, Supplementary Material]; the disentanglement windows are chosen individually~\cite{souza2001prb} for each compound, since a window adequate for one carbide can fail to reproduce the Fermi surface of another.

The electron energies, phonon frequencies, and electron--phonon matrix elements are then interpolated onto fine uniform grids. Because the scattering-in sum in Eq.~(\ref{eqn:ibte}) requires every final state $j\mathbf{k}'=i\mathbf{k}+\mathbf{q}$ to lie on the electron grid, the electron and phonon meshes are taken identical, $N^{3}_{\mathbf{k}}=N^{3}_{\mathbf{q}}$; this constraint makes the cost grow as $N^{6}$. Only states within $3$~eV of $\mu$ are retained, a window wide enough that the $(\varepsilon_{i\mathbf{k}}-\mu)^{2}$ weighting in Eq.~(\ref{eqn:K}), which peaks away from the chemical potential, is fully captured at the highest temperatures considered. The energy-conserving delta functions are evaluated with the adaptive smearing scheme~\cite{li2014cpc,li2015prb}, in which each transition is assigned a width from the local band and phonon group velocities rather than a single global broadening. Equation~(\ref{eqn:ibte}) is iterated from the SERTA solution with Broyden mixing (parameter $0.7$) to a maximum of $50$ iterations. Temperature enters only through $f_{i\mathbf{k}}$, and the phonon occupations and distinct temperatures do not couple in Eq.~(\ref{eqn:ibte}); each temperature is therefore an independent linear solve, and results computed at a subset of temperatures are exact rather than interpolated.

\bibliographystyle{elsarticle-num}
\bibliography{ref}

\end{document}